\documentclass[aps, llncs, preprint]{revtex4-1} 
\usepackage{graphicx}
\usepackage{amsmath}
\usepackage{float}
\usepackage{color}
\usepackage{subcaption}
\usepackage{amssymb}
\usepackage{braket}
\usepackage{rotating}
\usepackage{multirow}
\usepackage[bottom]{footmisc}
\usepackage{adjustbox}
\usepackage{caption}
\begin{document}
\title{Non-ballistic transport characteristics of superconducting point-contacts} 

\author{Ritesh Kumar}

\author{Goutam Sheet}
\email{goutam@iisermohali.ac.in}
\affiliation{Department of Physical Sciences, Indian Institute of Science Education and Research (IISER) Mohali, Sector 81, S. A. S. Nagar, Manauli, PO 140306, India}

 
\begin{abstract}

In the ``ballistic" regime, the transport across a normal metal (N)/superconductor (S) point-contact is dominated by a quantum process called Andreev reflection. Andreev reflection causes an enhancement of the conductance below the superconducting energy gap, and the ratio of the zero-bias and the high-bias conductance cannot be greater than 2 when the superconductor is conventional in nature. In this regime, the features associated with Andreev reflection also provide energy and momentum-resolved spectroscopic information about the superconducting phase. Here we theoretically consider various types of non-ballistic N/S point contacts within a network resistor model and show that even when the superconductor under investigation is simple conventional in nature, depending on the shape, size and anatomy of the point contacts, a wide variety of spectral features may appear in the conductance spectra. Such features may misleadingly mimic theoretically expected signatures of exotic physical phenomena like Klein tunneling in topological superconductors, Andreev bound states in unconventional superconductors, multiband superconductivity and Majorana zero modes.

\end{abstract}
\maketitle


Point contact spectroscopy\cite{Naidyuk} has been used by the low-temperature physics community as a powerful probe for investigating Fermi surface properties of complex materials for more than five decades\cite{Yanson,Duif,Naidyuk,Jansen}. One area of research where point contact spectroscopy has found tremendous success is superconductivity\cite{Blonder,DeWilde,Heil,Laube,Mao,Szabo,Walti,Yanson2}. When a sharp tip of a normal metal is brought in physical contact with a superconductor to form a superconducting point contact, upon the application of a voltage bias ($V$), the normal current (due to electrons) in the normal metal gets converted into a supercurrent (due to Cooper pairs) in the superconductor through a quantum process popularly known as Andreev reflection\cite{Andreev}. When $V$ across an N/S point contact is smaller than the energy scale of the superconducting energy gap ($\Delta/e$, $e$ being the charge of a single electron), the electrons in the metallic side, after passing through the point contact interface, fail to find single particle states in the superconducting side. These electrons reflect back (Andreev reflection) as holes in the metallic side leaving behind a Cooper pair to propagate inside the superconductor. Andreev reflection at superconducting point contacts have been used as a powerful spectroscopic tool for extracting energy, momentum and spin resolved spectroscopic information about the superconductors\cite{Naidyuk}. In such experiments, the current-voltage ($I-V$) characteristics of the point contacts are measured and the non-linearities in the characteristics, if any, are studied. Andreev reflection causes such non-linearity because as soon as $V = \Delta/e$, the conductance of the point contact gets enhanced due to Andreev reflection. However, one important aspect which is often ignored in such studies is that the intrinsic $I-V$ characteristics of superconductors are also non-linear. The effect of this non-linearity in a point contact spectroscopy experiment involving a superconducting point contact can be avoided only if the transport is allowed to take place in a transport regime where the bulk resistivities of the materials forming the point contact do not contribute to the overall point contact resistance. Fortunately, nano-positioning technology has empowered experimentalists to tune the size of the point contacts such that the effective contact diameter ($a$) can be made smaller than the elastic mean free path of the materials forming the point contact and make transport happen in the so-called ``ballistic regime" where the contact resistance is given by Sharvin's\cite{Sharvin} formula $R_S =\dfrac {2h}{e^2 (ak_F)^2}$, where $k_f$ is the Fermi momentum and $h$ is the Planck's constant. Clearly, in this regime, the bulk resistivity of the materials do not decide the point contact resistance. However, if the point contact is away from the ballistic regime, the so-called Maxwell's resistance\cite{Maxwell} ($R_M = \rho/2a$) also contributes to the over-all resistance. Here, the bulk resistivity $\rho$ cannot be ignored. In this paper we perform numerical simulations to show that depending on the way $R_M$ mixes with $R_S$ in a superconducting point contact which is away from the ballistic regime, one may obtain various types of spectra, some of which may mimic features associated with exotic unconventional nature of superconductivity or other relativistic effects.

\begin{figure}[h!]
		\includegraphics[width=.7\textwidth]{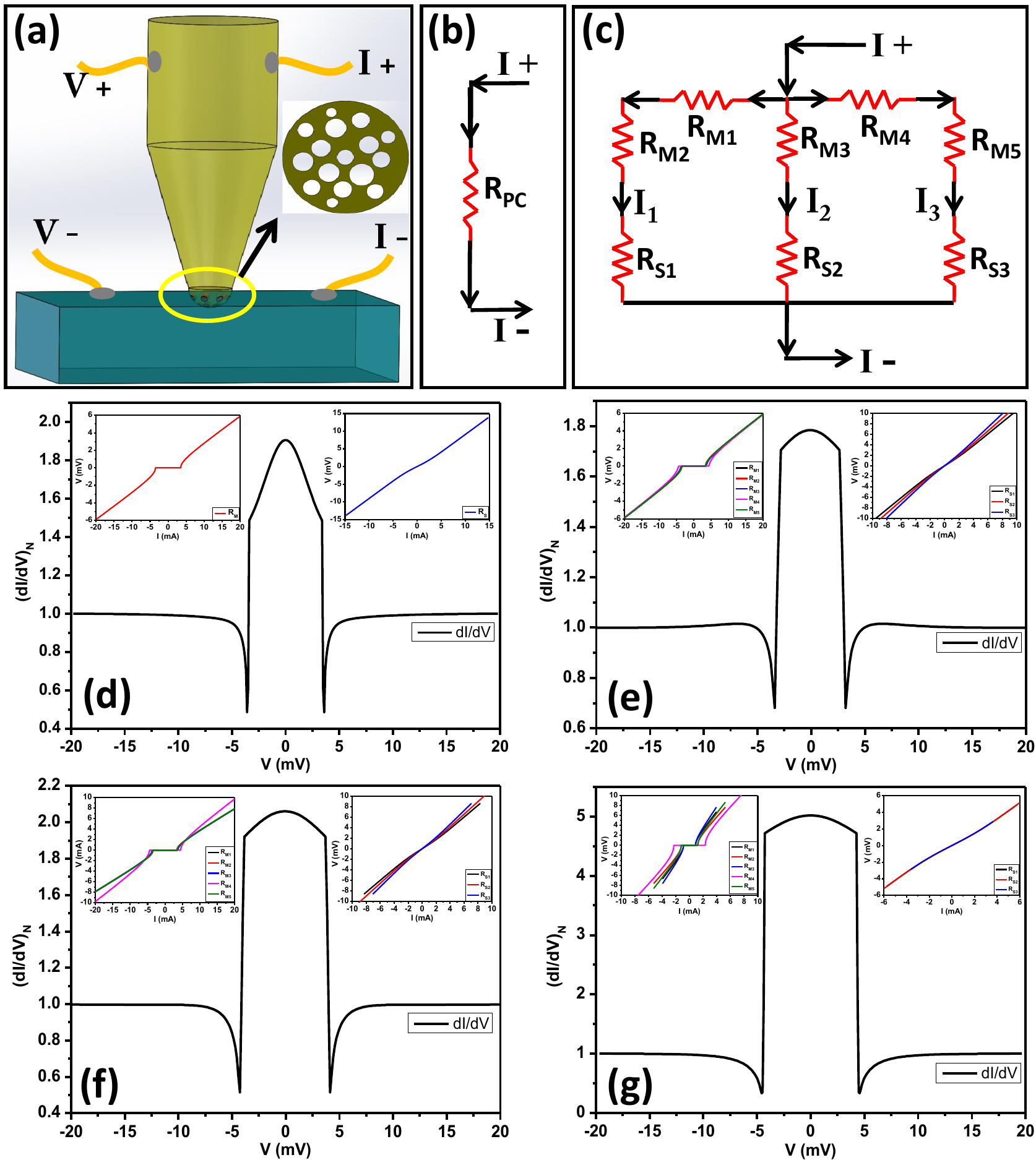}
		\caption{(a) Schematic of a point contact (PC). $Inset:$ Model of micro-constrictions within a contact. (b) $R_{PC}$ as a resistor for a single constriction. (c) Multiple constrictions under a point contact modelled as a resistor network where $R_{Mi}$ represent resistance due to Maxwell's contribution and $R_{Si}$ due to Sharvin's resistance (d) $dI/dV$ vs. $V$ spectrum for a single microconstriction as shown in (b). (e,f,g) $dI/dV vs. V$ spectra assuming multiple microconstrictions in parallel. The inset shows $I-V$ characteristics corresponding to the Maxwell's resistance ($R_M$) (left) and  Sharvin's resistance ($R_S$) (right). The parameters used to generate the spectra can be found in supplemental material (Table I).}	

\end{figure}

\begin{figure}[h!]
		\includegraphics[width=.7\textwidth]{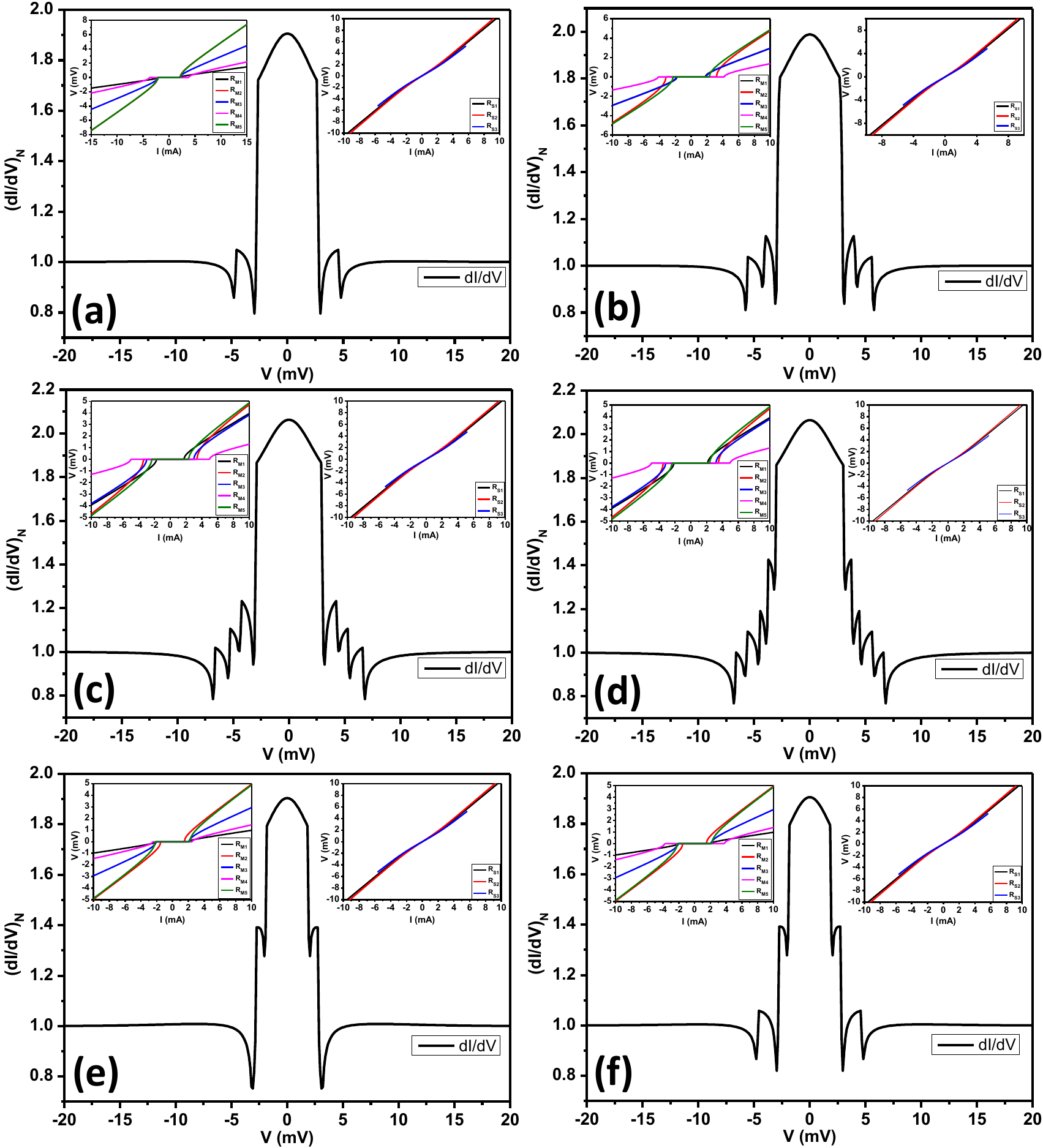}
		\caption{ $dI/dV$ vs. $V$ spectra with multiple conductance dips for the multi-constriction model shown in figure 1(c). (a),(b),(c) and (d) show emergence of 2, 3, 4 and 5 conductance dips respectively. (e) and (f) $dI/dV$ vs. $V$ spectra showing conductance dips at low energy. The inset shows $I-V$ characteristics corresponding to the Maxwell's resistance ($R_M$) (left) and  Sharvin's resistance ($R_S$) (right). The parameters used to generate the spectra can be found in supplemental material (Table II).}	

\end{figure}

The ballistic transport characteristics across a superconducting point contact are well understood within the one-dimensional model developed by Blonder-Tinkham and Klapwijk (BTK)\cite{Blonder}. In a conventional superconductor\cite{Tinkham}, the density of states (DOS) of the quasiparticles (single particle states) diverges at $E = eV = \pm\Delta$, when $V$ is the bias voltage applied across an N/S point contact, as DOS $N(E) = Re(\frac{E}{\sqrt{E^2-\Delta^2}})$.  For the typical conventional superconductors, $\Delta$ is found to be $\sim$ meV. In this energy scale, the DOS in the metallic side of an N/S point contact can be approximated to be constant with energy. BTK modelled the interface potential barrier as a one-dimensional delta function $V_0\delta(x)$. To characterize the strength of the potential barrier at the interface between the metal and the superconductor, BTK defined a dimensionless parameter $Z = \frac{V_0}{\hbar v_F}$, where $v_F$ is the Fermi velocity. $Z = 0$ means the barrier is transparent and in this extreme case, for $eV < \Delta$, the electrons undergo Andreev reflection with a probability $=$ 1. However, when $Z$ is non-zero, in addition to Andreev reflection the electrons can also undergo normal electrons and consequently, the Andreev reflection probability becomes less than 1. Assuming the electron and hole wave functions to be plane waves and solving the Bogoliubov-DeGennes equations, the Andreev reflection ($A(E)$) and the normal reflection ($B(E)$) probabilities were calculated by BTK. Then the current through the point contact could be calculated as $I_{NS} \propto N(0). v_F \int_{-\infty}^{+\infty} [f_0(E-eV)-f_0(E)][1+A(E)-B(E)] dE$\\
 where, $N(0)$ is the density of states at the Fermi level. This is then used to calculate the $dV/dI$ vs. $V$ spectra. Such theoretically generated spectra are then used to fit the experimental spectra to extract the value of the superconducting energy gap ($\Delta$). The Andreev reflection spectra calculated using BTK theory for different values of $Z$ are shown in the supplemental materials. For $Z= 0$, at very low temperatures, it is found that the spectrum is flat for $V \leq \Delta/e$. With increasing $Z$, two peaks at $V = \pm \Delta/e$ develop. This double peak structure in superconducting point contacts is a hallmark signature of Andreev reflection. When $Z$ is large, the peaks become very sharp and mimic the coherence peaks of the superconducting tunnelling spectra. Thus, local tunnelling spectroscopy can be thought of as a special case of point contact Andreev reflection spectroscopy. 
 

\begin{figure}[h!]
		\includegraphics[width=.7\textwidth]{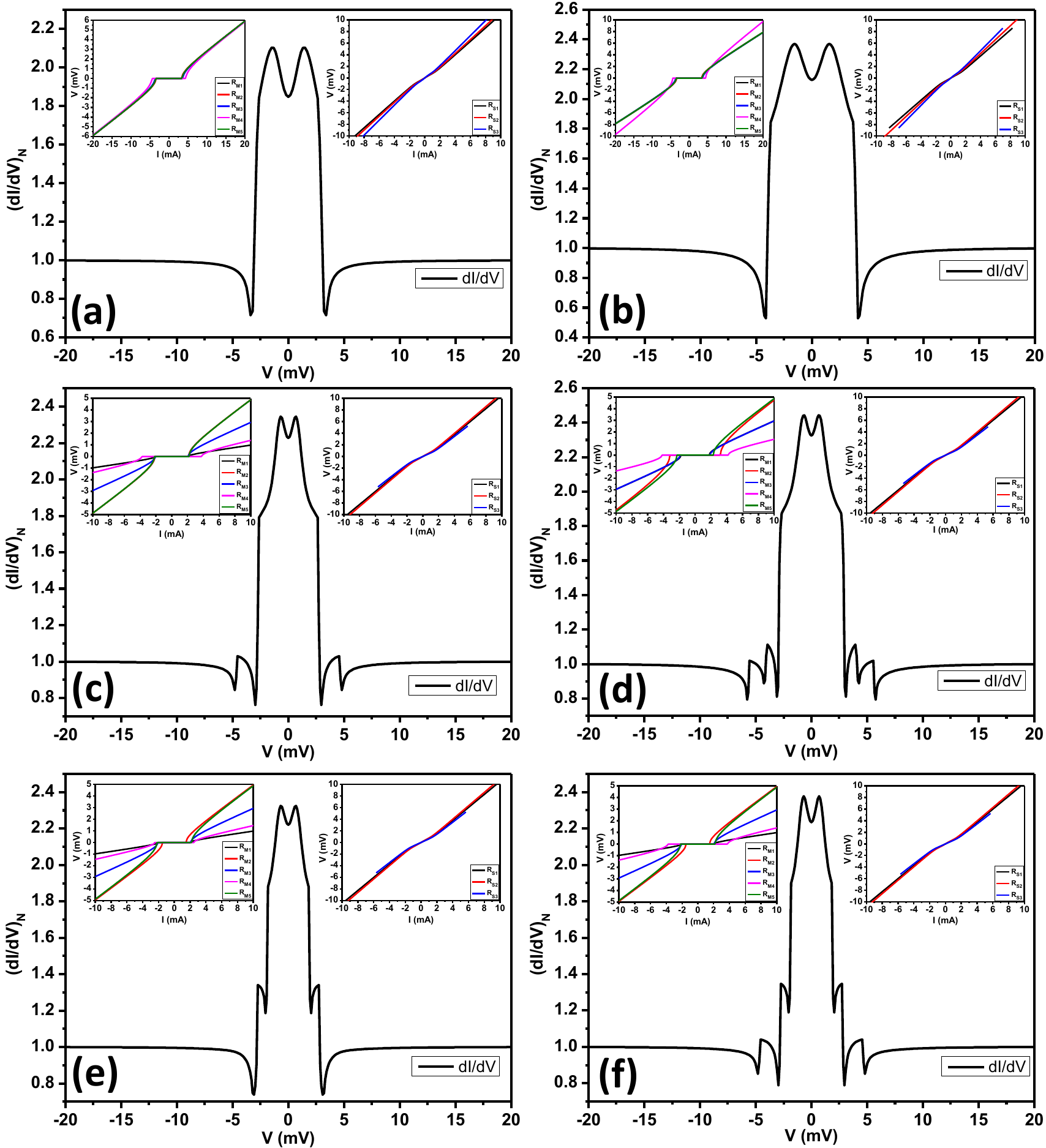}
		\caption{$dI/dV$ vs. $V$ spectra showing both Andreev reflection features and the signatures of critical current of micro-constrictions. The spectra (a,b) and (c,d,e,f) are obtained by increasing the percentage of $R_S$ component to the total resistance in the Figure 1(e,f) and 2(a,b,e,f) respectively. The parameters used to generate the spectra can be found in supplemental material (Table III).}	

\end{figure}

\begin{figure}[h!]
	\includegraphics[width=.7\textwidth]{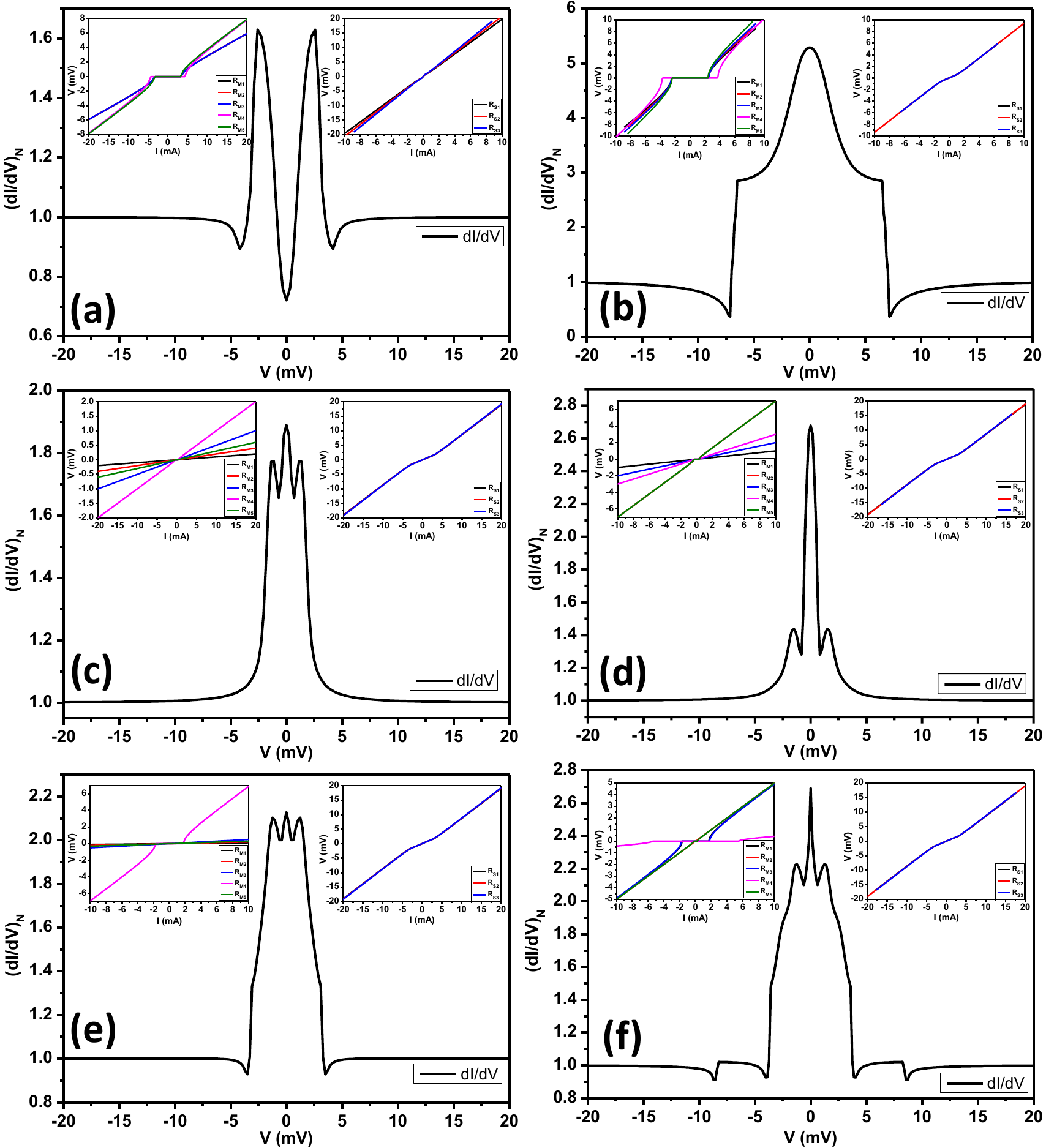}
	\caption{$dI/dV$ vs. $V$ spectra generated theoretically by adding $R_S$ and effective $R_M$ for a multi-constriction point contact in intermediate regime. (a) Spectrum with high $Z$ showing strong Andreev reflection along with the conductance dips. (b) Step-like feature seen in a conductance spectrum along with conductance dips. The zero-bias conductance peak (ZBCP) is due to low the critical current of certain constrictions. (c) Spectrum showing low critical current driven ZBCP but no conductance dips. (d) Spectrum obtained with increased contribution of $R_M$ in (c). (e) ZBCP appearing along with conductance dips due to co-existence of constrictions with high and low values of critical current. (f) Sharp ZBCP along with multiple conductance dips. The parameters used to generate the spectra can be found in supplemental material (Table IV).}	
	
\end{figure}


The discussion above is valid only when the point contacts are strictly in the ballistic regime of transport and $R_M$ has no role to play\cite{Sheet}. In order to understand how the mixing of $R_M$ will affect the shape of the point contact spectra for a superconducting point contact, we have considered the point contacts in an intermediate regime of transport where the total point contact resistance is given by Wexler's formula\cite{Wexler}: $R_{PC}=\frac{2h/e^2}{(ak_F)^2} + \Gamma(l/a)\frac{\rho(T)}{2a}$ where, $\Gamma(l/a)$ is a numerical factor close to unity, $a$ is the contact diameter and $2h/e^2$ is the fundamental quantum of resistance, $\rho(T)$ is the temperature dependent normal state resistivity of the superconducting material. Since the first term ($R_S$) is proportional to $1/a^2$ and the second term ($R_M$) goes as $1/a$, for smaller $a$, the first term dominates over the second term. This is when pure Andreev reflection is seen. When $a$ is larger, the contribution of the second term ($R_M$) cannot be ignored. 
In order to investigate such non-ballistic transport characteristics, we first need to look at the anatomy of a point contact. When a sharp metallic tip approaches a superconducting sample, first it forms a tunneling barrier. On further approaching, microscopic cracks (micro-constrictions) develop at the contact region due to which the tunneling barrier develops ``shorts" and electronic transport in the so-called metallic regime can happen through these micro-constrictions (see Figure 1(a)). In an ideal case scenario, a point contact may be thought of as a single micro-constriction (Figure 1(b)). When there are multiple micro-constrictions, each one of them can be of different size and shape, and the total current injected through the point contact will be divided into them as if they are some resistors connected in a parallel configuration ($inset$ of Figure 1(c)). In a real experiment, it may not be possible to precisely determine the shape and size of such individual micro-constrictions. However, each micro-constriction can have distinct $I-V$ characteristics and depending on the way they mix with each other and contribute to the total current, the shape of the resulting $dI/dV$ vs. $V$ spectrum can be dramatically different. To investigate this effect numerically, we have modelled a point contact as a set of five micro-constrictions each having its own characteristic $R_M$ and $R_S$ which are connected with each other as a resistance network circuit as shown in Figure 1(c). $R_{M1}$ and $R_{M4}$ represent two purely thermal contacts which connect the constrictions 1, 2 and 3. We have generated the $I-V$ characteristics\cite{Gallop,comment} corresponding to each $R_M$ using the formula $V_\alpha = R_{M\alpha}I_{c\alpha}\sqrt{(I_i/I_{c\alpha})^2-1}$, where total current $I = \sum_{i}I_i$. $\alpha$ takes values 1,2,3,4,5. The index $i$ represents $i$-th micro-constriction ($i$ = 1,2,3) in the intermediate regime and $I_{c\alpha}$ is the critical current of the corresponding $R_{M\alpha}$. This formula is used to calculate $I_i$ vs. $V$ for all values of $I_i >I_{ci}$. The calculated $I-V$ thus obtained has been folded in order to obtain the characteristics for the current flowing in the opposite direction as well. The $I-V$ characteristics corresponding to $R_S$ have been obtained using the BTK theory for different constrictions keeping the gap ($\Delta$) same for all the constrictions (as it is the same superconductor for all). Then, we calculated the total $dV/dI$ vs. $V$ curves adding the $dV/dI$ characteristics corresponding to different $R_M$ and $R_S$ for the three arms of the network circuit (characterized by $I_1$, $I_2$ and $I_3$). After that, we found $dI/dV$ for the arms. The $dI/dV$ corresponding to the three arms were added to compute the final $dI/dV$ vs. $V$ graphs presented in the paper. Now, we have simply varied the $R_M$ and the $R_S$ corresponding to the micro-constrictions to generate a series of spectra and found that for certain combination of these parameters special spectral features may appear in conventional N/S point contacts that may closely resemble some exotic effects. We discuss our findings below. All the $I-V$ characteristics corresponding to respective $R_M$ and $R_S$ that we have considered are shown in the $inset$ of the figure panels. The details of the parameters used for generating the $I-V$ characteristics and the $dI/dV$ spectra are provided in the supplemental material.

In Figure 1(d), we show a representative $dI/dV$ vs. $V$ spectrum corresponding to $R_M$, where only one micro-constriction with a definite value of $I_c$ has been assumed (as in Figure 1(b)). Here, $I-V$ due to $R_M$ has been added to $I-V$ corresponding to $R_S$ to account for the transport in intermediate regime, as per Wexler's formula. The main characteristics of the $dI/dV$ vs. $V$ spectrum in Figure 1(d) are the two sharp conductance dips. Clearly, these dips are due to the non-linearities originating from the critical current (see the red curve in the $inset$). In practice, multiple micro-constrictions appear under a point contact\cite{Das}. In such cases, the contribution of $R_M$ from a large number of different constrictions can be added up in such a way that only one pair of conductance dips appear in the spectrum, but the dips may get broadened. In experiments, such broadened dip structures are often observed. Since in this case the low-bias conductance enhancement is not due to Andreev reflection, there is no upper limit of the ratio between the zero-bias conductance and the high-bias conductance. Under certain circumstances, as shown in Figure 1(f), the zero-bias conductance can even get exactly doubled thereby mimicking ``perfect" Andreev reflection as is expected due to Klein tunneling in topological superconductors\cite{Chang,Lee,Yu}. As shown in Figure 1(g), the zero-bias conductance can even become far greater than 2 and may look like a signature of robust zero-energy bound states\cite{Yu}. In all such cases, the position of the dips can appear in arbitrary places in the energy axis, depending on the choice of $I-V$ corresponding to $R_{M\alpha}$.

All the spectra that we discussed above displayed only one pair of conductance dips. By changing the proportion in which the critical current driven $I-V$ characteristics of different micro-constrictions mix with each other, we have found a number of spectra where multiple conductance dips were also observed. In Figure 2(a,b,c,d), we show such representative spectra with 2, 3, 4 and 5 conductance dips obtained only by assuming different characteristics of five micro-constrictions all of which together define the anatomy of a given point contact. Such multiple dips\cite{Kayyalha}, when appear in experiments, are often related to multiple Andreev reflections\cite{Rafael}, topological superconductivity\cite{Wang}, etc. The spectra in Figure 2(c) and Figure 2(d) can be erroneously taken as signature of quantum oscillations (or, simply some quantum quantization effect) if the role of $R_M$ is not considered. Furthermore, as shown in Figure 2(e) and Figure 2(f), the $I_c$ driven dips may also appear at very low energies as tiny downward kinks in the conductance spectrum -- in the past such features were attributed to multi-gap superconductivity\cite{Rourke,Sheet1,Park}.

Now, let us consider the point contacts in a non-ballistic regime where the contribution of $R_S$ is also prominent and $R_M$ and $R_S$ are comparable. This can happen in two ways. Either all of the micro-constrictions under a point-contact reach a regime where $R_S$ and $R_M$ for each is in the ``intermediate" regime. Or, some of the micro-constrictions remain closer to the thermal regime while others are in the ballistic regime of transport. In the most simple model, the contribution of $R_S$ and $R_M$ can be simply added as is done in Wexler's formula. In this case, the $I-V$ characteristics are expected to be non-linear due to critical current as well as due to Andreev reflection. In Figure 3 (a-f) we show the appearance of the double-peak feature associated with Andreev reflection, when the percentage of the $R_S$ component to the respective total resistance as in Figure 2 has been increased. The conductance dips due to the critical current dominated $I-V$ characteristics corresponding to $R_M$ coexist with the Andreev reflection features. In the literature, such spectra are ubiquitously seen and are often attributed to more complex physical phenomena like Josephson tunneling\cite{Shan}, superconducting proximity effect\cite{Srikanth,Son}, spin-triplet superconductivity\cite{Laube,Andrew} etc.

Finally, we also explored other possible ways in which the point contact Andreev reflection spectra can deviate from a canonical shape and found a number of spectra with unique features which might falsely indicate the emergence of novel fundamental excitations/modes in a solid. A collection of such spectra along with the assumed $I-V$ characteristics corresponding to $R_M$ and $R_S$ are shown in Figure 4. The spectrum shown in Figure 4(a) has conductance dips and very sharp and prominent Andreev reflection features. In Figure 4(b), we show the emergence of a large zero-bias conductance peak. This large zero-bias conductance enhancement, in addition to the Andreev reflection features, is due to some of the non-ballistic constrictions having low $I_c$. In a real experiment such a feature may appear in a robust way and mimic the Andreev bound states as expected in point contacts with certain special unconventional superconductors. As shown in Figure 4(c,d,e,f), when for some of the micro-constrictions exhibit extremely low $I_c$, depending on the distribution of $I_c$ in different micro-constrictions, the shape and size of the zero-bias conductance peak (ZBCP) can vary. In some extreme cases, the ZBCP can be very sharp and erroneously hint to the possibility of exotic pairing in the superconductor being investigated\cite{Yu}. It is also understood that such features will also evolve with changing temperatures and magnetic fields as the critical current for each constriction also changes with temperature and magnetic fields.

In conclusion, in order to understand the origin of special spectral features that ubiquitously appear in point contact Andreev reflection spectroscopy, we have theoretically modelled a superconducting point-contact as an ensemble of a number of micro-constrictions through which electronic transport takes place. We have considered the shape and size of the micro-constrictions to be different and consequently giving rise to different shape of the $I-V$ characteristics. We have then mixed such $I-V$ characteristics in different proportions and found wide variety of spectra appear closely resembling the features expected for certain complex physical phenomena. We have also allowed some of the micro-constrictions to be in the ballistic regime and found that depending on the way Maxwell's ($R_M$) and Sharvin's ($R_S$) resistance mix with each other, spectral features mimicking even more complex processes may appear. Our model calculations can be used to analyze such non-ballistic spectra and such analysis can help extract spectroscopic information selectively from the ballistic micro-constrictions in the ``intermediate" regime point contacts. Therefore, the exotic features appearing in Andreev reflection spectroscopy can be attributed to exotic physics only after carefully ruling out the contribution of the non-ballistic characteristics discussed here.

The authors thank Mona Garg for her help. GS acknowledges financial support from Swarnajayanti fellowship awarded by the Department of Science and Technology (DST), Govt. of India (grant number DST/SJF/PSA-01/2015-16).

\bibliography{References}

\begin{thebibliography}{100}
\bibitem{comment} In fact, as long as the shape of the $I-V$ curve corresponding to $R_M$ shows the dissipationless state accurately, the exact formula that gives rise to the $I-V$ characteristics will not make any qualitative difference in the shape of the calculated characteristics.
\end{thebibliography}

\newpage

\begin{center}
	
	\textbf{Supplemental materials}
	
\end{center}
\underline{\textit{\textbf{Point Contact Andreev Reflection Spectroscopy \& BTK calculations:}}}

In figure 1(a), we show the schematic of a typical point contact setup with electrical connections where a normal metal (N) tip gently touches a superconducting sample (S). Figure 1(b) depicts the Andreev reflection process at an N/S point contact interface. The right hand side shows the density of states (DOS) profile of a conventional superconductor. This is directly obtained from Bardeen-Cooper-Schriffer (BCS) formalism, where DOS is given by $N(E) = Re(\frac{E}{\sqrt{E^2-\Delta^2}})$. $\Delta$ is the superconducting energy gap ($\sim meV$ for a conventional superconductor) -- in this scale, the DOS in the metal remain independent of energy. The flat density of states for the metal is shown in the left hand side. A delta function potential barier at the interface is assumed in the BTK theory. The calculated differential conductance spectra within the BTK formalism for low to high barriers at $T = 1.6K$ are shown in figure 1(c,d,e,f).   
\begin{figure}[h!]
	\centering
	\includegraphics[width=.45\textwidth]{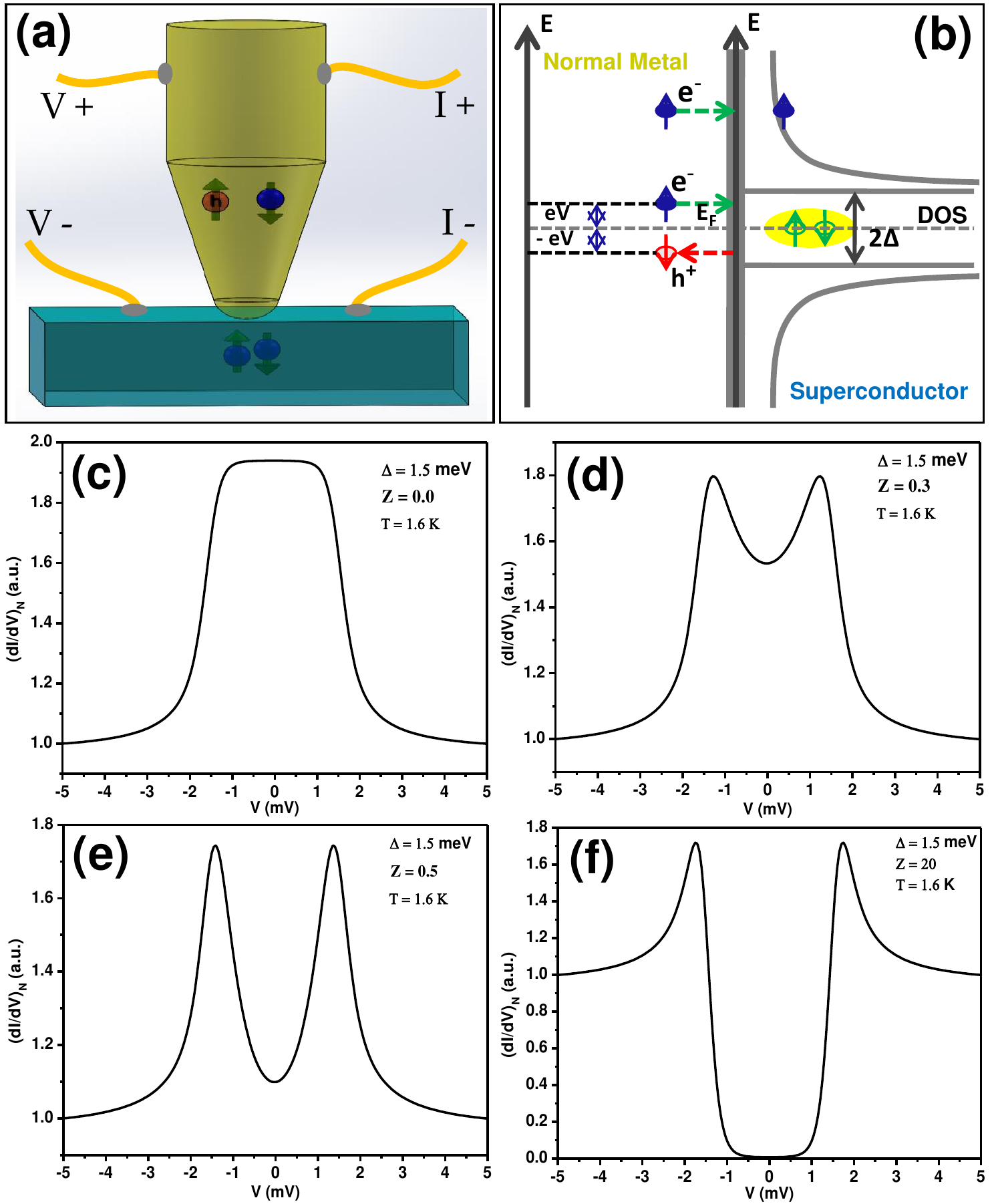}
	\caption{(a) Schematic of point contact setup with electrical connections. (b) Schematic depicting the Andreev reflection process. (c),(d),(e) and (f) Normalized differential conductance spectra calculated within standard BTK model for different values of $Z$.}	
	
\end{figure}\\\\

\underline{\textit{\textbf{Details of Maxwell's Resistance ($R_M$):}}}

When a point contact is away from the ballistic regime, as per Wexler's formula, the Sharvin's
resistance ($R_S$) and the Maxwell's resistance ($R_M$) simply add up (modulo a coefficient of $R_M$) as if they are in series. In order to understand how they will mix with each other in a non-ballistic point contact, one first needs to look at the experimental method.
The quantity that is plotted is differential conductance ($dI/dV$) with $V$. For this, a dc current, mixed with a sine wave (of very small amplitude compared to the magnitude of the dc component) is passed through the point contact. The dc voltage drop across the point contact becomes the horizontal ($V$) axis of the final plot and the ac drop (typically measured by a lock-in amplifier) is proportional to the vertical ($dI/dV$) axis. The dc voltage drop is given by $V = I( R_S + R_M)$. Now, when $I$ just crosses the critical current $I_c$, the value of $R_M$ suddenly becomes non-zero (from zero in the dissipation-less state). Obviously, this change will appear in the plotted data at certain values of $V$.
\begin{figure}[h!]
	\centering
	\includegraphics[width=.4\textwidth]{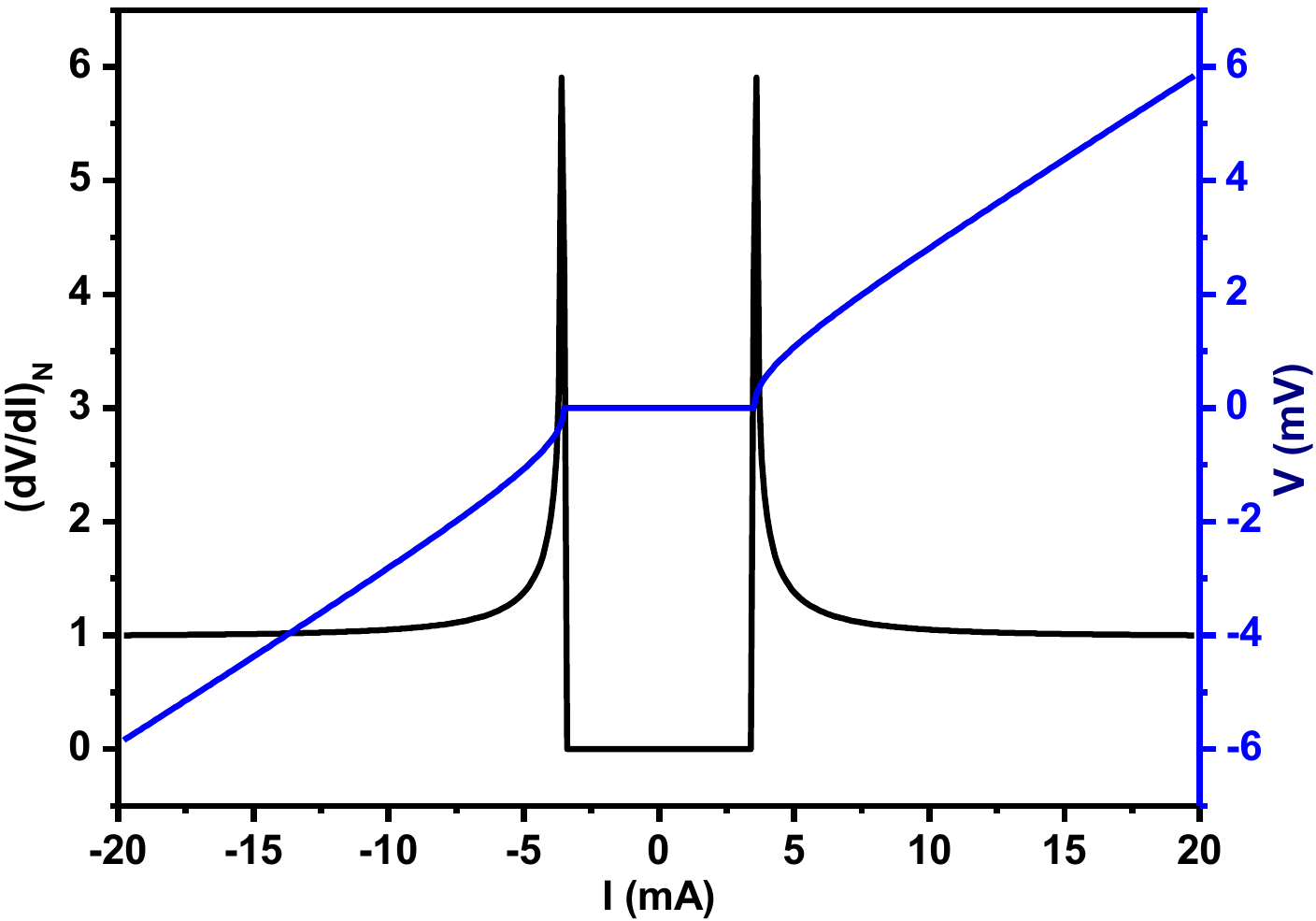}
	\caption{Typical $I-V$ characteristics corresponding to $R_M$ and its derivative ($dV/dI$).}	
\end{figure}
Again, in the final plot, the horizontal axis will need to be multiplied by the total point contact resistance in order to make it $V$. Now, these peaks in $dV/dI$ will appear as dips in $dI/dV$. When the $R_S$ component (and any other component depending on the nature of the electrodes forming the point contact) is also added, the $dV/dI$ below $V = I_c(R_M + R_S)$ will be non-zero. This is how the critical current enters the picture! For multiple constrictions with different critical currents, the resultant spectra may show multiple conductance dips (and the other features as discussed in the manuscript).\\
To note, in the point contacts away from the ballistic regime, the horizontal ($V$) axis cannot be strictly taken as the energy axis as the dissipative processes are allowed in such regimes of transport.\\
\\
\begin{table}
	\begin{center}
		\tiny
		\caption{\textbf{The parameters used to generate $V-I$ characteristics in figure 1(d-g) in the manuscript. The values of $R_M$ components and the respective critical current ($I_C (mA)$ ) for the multi-constriction model (shown in figure 1(c)) are listed in ``Details of Maxwell's Resistance ($R_M$)". The parameters to generate $I-V$ characteristics (from BTK theory) corresponding to $R_S$ with weightage factor are listed in ``Details of Sharvin's Resistance ($R_S$)".}}
		\label{tab:table1}
		\begin{tabular}{|p{1.2cm}||p{1.2cm}|p{1.2cm}|p{1.2cm}|p{1.2cm}|p{1.2cm}|p{1.2cm}|p{1.2cm}|p{1.2cm}|p{1.2cm}|p{1.2cm}|}
			
			\hline
			\multicolumn{11}{|c|}{\textbf{Details of Maxwell's Resistance ($R_M$)}}\\
			\hline
			
			\multirow{2}{*}{\textbf{(d)}}&\multicolumn{5}{c|}{$R_{M}$} & 
			\multicolumn{5}{c|}{$I_{C} $}\\
			\cline{2-11}
			&\multicolumn{5}{c|}{0.3} & \multicolumn{5}{c|}{3.5}\\ 
			\cline{1-11}
			
			&$R_{M1}$ &$I_{C1}$&$R_{M2}$ &$I_{C2}$&$R_{M3}$ &$I_{C3}$ &$R_{M4}$ &$I_{C4}$ &$R_{M5}$&$I_{C5}$\\
			\cline{1-11}
			\multirow{1}{*}{\textbf{(e)}}&0.3 &3.6 &0.3 &3.7 &0.5 &4.5 &0.4 &3.4 &0.4&3.5\\ 
			\cline{1-11}
			
			\multirow{1}{*}{\textbf{(f)}}&0.4 &3.6 &0.4 &3.7 &0.5 &4.5 &0.4 &3.4 &0.4&3.5\\ 
			\cline{1-11}
			
			\multirow{1}{*}{\textbf{(g)}}&1.75&0.875 &2 &0.9 &1.4 &2.3 &1.5 &1.2 &1.7&1.225\\ 
			\cline{1-11}
			
			\multicolumn{11}{|c|}{\textbf{Details of Sharvin's Resistance ($R_S$)}}\\
			\hline
			&&\multicolumn{2}{c|}{weight factor}&\multicolumn{2}{c|}{$ \Delta (meV) $} &\multicolumn{2}{c|}{$ \Gamma (meV) $}  &\multicolumn{2}{c|}{$ Z $}  &$ T (K) $\\
			\cline{1-2}\cline{3-4}\cline{5-6}\cline{7-8}\cline{9-11}
			
			\multirow{1}{*}{\textbf{(d)}}&$R_{S}$&\multicolumn{2}{c|}{0.7}&\multicolumn{2}{c|}{0.7} &\multicolumn{2}{c|}{0.001}  &\multicolumn{2}{c|}{0.001} & 10\\
			\cline{1-2}\cline{3-4}\cline{5-6}\cline{7-8}\cline{9-11}

			\multirow{3}{*}{\textbf{(e)}}&$R_{S1}$&\multicolumn{2}{c|}{0.3}&\multicolumn{2}{c|}{0.7}&\multicolumn{2}{c|}{0.001}  &\multicolumn{2}{c|}{0.35}&10\\
			\cline{2-3}\cline{4-5}\cline{5-6}\cline{7-8}\cline{9-11}
			&$R_{S2}$&\multicolumn{2}{c|}{0.35}&\multicolumn{2}{c|}{0.7} &\multicolumn{2}{c|}{0.001}&\multicolumn{2}{c|}{0.45}&10\\
			\cline{2-3}\cline{4-5}\cline{5-6}\cline{7-8}\cline{9-11}
			&$R_{S3}$&\multicolumn{2}{c|}{0.3}&\multicolumn{2}{c|}{0.7} &\multicolumn{2}{c|}{0.001}&\multicolumn{2}{c|}{0.55}&10\\
			\cline{1-2}\cline{3-4}\cline{5-6}\cline{7-8}\cline{9-11}
			
			\multirow{3}{*}{\textbf{(f)}}&$R_{S1}$&\multicolumn{2}{c|}{0.35}&\multicolumn{2}{c|}{0.7} &\multicolumn{2}{c|}{0.001}  &\multicolumn{2}{c|}{0.35}&10\\
			\cline{2-3}\cline{4-5}\cline{5-6}\cline{7-8}\cline{9-11}
			&$R_{S2}$&\multicolumn{2}{c|}{0.4}&\multicolumn{2}{c|}{0.7} &\multicolumn{2}{c|}{0.001}&\multicolumn{2}{c|}{0.45}&10\\
			\cline{2-3}\cline{4-5}\cline{5-6}\cline{7-8}\cline{9-11}
			&$R_{S3}$&\multicolumn{2}{c|}{0.35}&\multicolumn{2}{c|}{0.7} &\multicolumn{2}{c|}{0.001}&\multicolumn{2}{c|}{0.55}&10\\
			\cline{1-2}\cline{3-4}\cline{5-6}\cline{7-8}\cline{9-11}
			
			\multirow{3}{*}{\textbf{(g)}}&$R_{S1}$&\multicolumn{2}{c|}{1.25}&\multicolumn{2}{c|}{0.7} &\multicolumn{2}{c|}{0.001}  &\multicolumn{2}{c|}{0.05} & 10 \\
			\cline{2-3}\cline{4-5}\cline{5-6}\cline{7-8}\cline{9-11}
			&$R_{S2}$&\multicolumn{2}{c|}{0.5}&\multicolumn{2}{c|}{0.7} &\multicolumn{2}{c|}{0.001}&\multicolumn{2}{c|}{0.07}&10 \\
			\cline{2-3}\cline{4-5}\cline{5-6}\cline{7-8}\cline{9-11}
			&$R_{S3}$&\multicolumn{2}{c|}{0.55}&\multicolumn{2}{c|}{0.7} &\multicolumn{2}{c|}{0.001}&\multicolumn{2}{c|}{0.09}&10 \\
			\cline{1-2}\cline{3-4}\cline{5-6}\cline{7-8}\cline{9-11}
		\end{tabular}
	\end{center}
\end{table}

\begin{table}
	\begin{center}
		\tiny
		\caption{\textbf{The parameters used to generate $V-I$ characteristics in figure 2(a-g) in the manuscript. The values of $R_M$ components and the respective critical current ($I_C (mA)$ ) for the multi-constriction model (shown in figure 1(c)) are listed in ``Details of Maxwell's Resistance ($R_M$)''. The parameters to generate $I-V$ characteristics (from BTK theory) corresponding to $R_S$ with weightage factor are listed in ``Details of Sharvin's Resistance ($R_S$)".}}
		\label{tab:table1}
		\begin{tabular}{|p{1.2cm}||p{1.2cm}|p{1.2cm}|p{1.2cm}|p{1.2cm}|p{1.2cm}|p{1.2cm}|p{1.2cm}|p{1.2cm}|p{1.2cm}|p{1.2cm}|}
			
			\hline
			\multicolumn{11}{|c|}{\textbf{Details of Maxwell's Resistance ($R_M$)}}\\
			\hline
			
			&$R_{M1}$ &$I_{C1}$&$R_{M2}$ &$I_{C2}$&$R_{M3}$ &$I_{C3}$ &$R_{M4}$ &$I_{C4}$ &$R_{M5}$&$I_{C5}$\\
			\cline{1-11}
			
			\multirow{1}{*}{\textbf{(a)}}&0.1&2.05&0.3&2.1&0.15&3.8&0.5&2.15&0.5&2.2\\ 
			\cline{1-11}
			
			\multirow{1}{*}{\textbf{(b)}}&0.3 &1.8 &0.3&1.9&0.15&4.2&0.5 &2.3 &0.5&3.2\\ 
			\cline{1-11}
			
			\multirow{1}{*}{\textbf{(c)}}&0.4&1.75&0.4&3&0.15&5&0.5&2.35&0.5&3.4\\ 
			\cline{1-11}
			
			\multirow{1}{*}{\textbf{(d)}}&0.4&2.15&0.4&3.2&0.15&5&0.5&2.4&0.5&3.5\\ 
			\cline{1-11}
			
			\multirow{1}{*}{\textbf{(e)}}&0.1&2.05&0.3&2.1&0.15&2.5&0.5 &1.5 &0.5&2.2\\
			\cline{1-11}
			
			\multirow{1}{*}{\textbf{(f)}}&0.1&2.05&0.3&2.1&0.15&3.8&0.5&1.5&0.5&2.2\\  
			\cline{1-11}
			
			\multicolumn{11}{|c|}{\textbf{Details of Sharvin's Resistance ($R_S$)}}\\
			\hline
			&&\multicolumn{2}{c|}{weight factor}&\multicolumn{2}{c|}{$ \Delta (meV) $} &\multicolumn{2}{c|}{$ \Gamma (meV) $}  &\multicolumn{2}{c|}{$ Z $}  &$ T (K) $\\
			\cline{1-2}\cline{3-4}\cline{5-6}\cline{7-8}\cline{9-11}
			
			\multirow{3}{*}{\textbf{(a)}}&$R_{S1}$&\multicolumn{2}{c|}{1.0}&\multicolumn{2}{c|}{0.7} &\multicolumn{2}{c|}{0.001}  &\multicolumn{2}{c|}{0.35} & 10 \\
			\cline{2-3}\cline{4-5}\cline{5-6}\cline{7-8}\cline{9-11}
			&$R_{S2}$&\multicolumn{2}{c|}{1.1}&\multicolumn{2}{c|}{0.7} &\multicolumn{2}{c|}{0.001}  &\multicolumn{2}{c|}{0.4} & 10 \\
			\cline{2-3}\cline{4-5}\cline{5-6}\cline{7-8}\cline{9-11}
			&$R_{S3}$&\multicolumn{2}{c|}{0.3}&\multicolumn{2}{c|}{0.7} &\multicolumn{2}{c|}{0.001}  &\multicolumn{2}{c|}{0.25} & 10 \\
			\cline{1-2}\cline{3-4}\cline{5-6}\cline{7-8}\cline{9-11}

			\multirow{3}{*}{\textbf{(b)}}&$R_{S1}$&\multicolumn{2}{c|}{1.0}&\multicolumn{2}{c|}{0.7} &\multicolumn{2}{c|}{0.001}  &\multicolumn{2}{c|}{0.35} & 10 \\
			\cline{2-3}\cline{4-5}\cline{5-6}\cline{7-8}\cline{9-11}
			&$R_{S2}$&\multicolumn{2}{c|}{1.2}&\multicolumn{2}{c|}{0.7} &\multicolumn{2}{c|}{0.001}  &\multicolumn{2}{c|}{0.4} & 10 \\
			\cline{2-3}\cline{4-5}\cline{5-6}\cline{7-8}\cline{9-11}
			&$R_{S3}$&\multicolumn{2}{c|}{0.3}&\multicolumn{2}{c|}{0.7} &\multicolumn{2}{c|}{0.001}  &\multicolumn{2}{c|}{0.25} & 10 \\
			\cline{1-2}\cline{3-4}\cline{5-6}\cline{7-8}\cline{9-11}
			
			\multirow{3}{*}{\textbf{(c)}}&$R_{S1}$&\multicolumn{2}{c|}{1.0}&\multicolumn{2}{c|}{0.7} &\multicolumn{2}{c|}{0.001}  &\multicolumn{2}{c|}{0.35} & 10 \\
			\cline{2-3}\cline{4-5}\cline{5-6}\cline{7-8}\cline{9-11}
			&$R_{S2}$&\multicolumn{2}{c|}{1.2}&\multicolumn{2}{c|}{0.7} &\multicolumn{2}{c|}{0.001}  &\multicolumn{2}{c|}{0.4} & 10 \\
			\cline{2-3}\cline{4-5}\cline{5-6}\cline{7-8}\cline{9-11}
			&$R_{S3}$&\multicolumn{2}{c|}{0.3}&\multicolumn{2}{c|}{0.7} &\multicolumn{2}{c|}{0.001}  &\multicolumn{2}{c|}{0.25} & 10 \\
			\cline{1-2}\cline{3-4}\cline{5-6}\cline{7-8}\cline{9-11}
			
			\multirow{3}{*}{\textbf{(d)}}&$R_{S1}$&\multicolumn{2}{c|}{1.0}&\multicolumn{2}{c|}{0.7} &\multicolumn{2}{c|}{0.001}  &\multicolumn{2}{c|}{0.35} & 10 \\
			\cline{2-3}\cline{4-5}\cline{5-6}\cline{7-8}\cline{9-11}
			&$R_{S2}$&\multicolumn{2}{c|}{1.2}&\multicolumn{2}{c|}{0.7} &\multicolumn{2}{c|}{0.001}  &\multicolumn{2}{c|}{0.4} & 10 \\
			\cline{2-3}\cline{4-5}\cline{5-6}\cline{7-8}\cline{9-11}
			&$R_{S3}$&\multicolumn{2}{c|}{0.3}&\multicolumn{2}{c|}{0.7} &\multicolumn{2}{c|}{0.001}  &\multicolumn{2}{c|}{0.25} & 10 \\
			\cline{1-2}\cline{3-4}\cline{5-6}\cline{7-8}\cline{9-11}
			
			\multirow{3}{*}{\textbf{(e)}}&$R_{S1}$&\multicolumn{2}{c|}{1.0}&\multicolumn{2}{c|}{0.7} &\multicolumn{2}{c|}{0.001}  &\multicolumn{2}{c|}{0.35} & 10 \\
			\cline{2-3}\cline{4-5}\cline{5-6}\cline{7-8}\cline{9-11}
			&$R_{S2}$&\multicolumn{2}{c|}{1.1}&\multicolumn{2}{c|}{0.7} &\multicolumn{2}{c|}{0.001}  &\multicolumn{2}{c|}{0.4} & 10 \\
			\cline{2-3}\cline{4-5}\cline{5-6}\cline{7-8}\cline{9-11}
			&$R_{S3}$&\multicolumn{2}{c|}{0.3}&\multicolumn{2}{c|}{0.7} &\multicolumn{2}{c|}{0.001}  &\multicolumn{2}{c|}{0.025} & 10 \\
			\cline{1-2}\cline{3-4}\cline{5-6}\cline{7-8}\cline{9-11}
			
			\multirow{3}{*}{\textbf{(f)}}&$R_{S1}$&\multicolumn{2}{c|}{1.0}&\multicolumn{2}{c|}{0.7} &\multicolumn{2}{c|}{0.001}  &\multicolumn{2}{c|}{0.35} & 10 \\
			\cline{2-3}\cline{4-5}\cline{5-6}\cline{7-8}\cline{9-11}
			&$R_{S2}$&\multicolumn{2}{c|}{1.1}&\multicolumn{2}{c|}{0.7} &\multicolumn{2}{c|}{0.001}  &\multicolumn{2}{c|}{0.4} & 10 \\
			\cline{2-3}\cline{4-5}\cline{5-6}\cline{7-8}\cline{9-11}
			&$R_{S3}$&\multicolumn{2}{c|}{0.3}&\multicolumn{2}{c|}{0.7} &\multicolumn{2}{c|}{0.001}  &\multicolumn{2}{c|}{0.25} & 10 \\
			\cline{1-2}\cline{3-4}\cline{5-6}\cline{7-8}\cline{9-11}
		\end{tabular}
	\end{center}
\end{table}

\begin{table}
	\begin{center}
		\tiny
		\caption{\textbf{The parameters used to generate $V-I$ characteristics in figure 3(a-g) in the manuscript. The values of $R_M$ components and the respective critical current ($I_C (mA)$ ) for the multi-constriction model (shown in figure 1(c)) are listed in ``Details of Maxwell's Resistance ($R_M$)". The parameters to generate $I-V$ characteristics (from BTK theory) corresponding to $R_S$ with weightage factor are listed in ``Details of Sharvin's Resistance ($R_S$)".}}
		\label{tab:table1}
		\begin{tabular}{|p{1.2cm}||p{1.2cm}|p{1.2cm}|p{1.2cm}|p{1.2cm}|p{1.2cm}|p{1.2cm}|p{1.2cm}|p{1.2cm}|p{1.2cm}|p{1.2cm}|}
			
			\hline
			\multicolumn{11}{|c|}{\textbf{Details of Maxwell's Resistance ($R_M$)}}\\
			\hline
			
			&$R_{M1}$ &$I_{C1}$&$R_{M2}$ &$I_{C2}$&$R_{M3}$ &$I_{C3}$ &$R_{M4}$ &$I_{C4}$ &$R_{M5}$&$I_{C5}$\\
			\cline{1-11}
			\multirow{1}{*}{\textbf{(a)}}&0.3 &3.6 &0.3 &3.7 &0.5 &4.5 &0.4 &3.4 &0.4&3.5\\ 
			\cline{1-11}
			
			\multirow{1}{*}{\textbf{(b)}}&0.4 &3.6 &0.4 &3.7 &0.5 &4.5 &0.4 &3.4 &0.4&3.5\\ 
			\cline{1-11}
			
			\multirow{1}{*}{\textbf{(c)}}&0.1&2.05&0.3&2.1&0.15&3.8&0.5&2.15&0.5&2.2\\ 
			\cline{1-11}
			
			\multirow{1}{*}{\textbf{(d)}}&0.3 &1.8 &0.3&1.9&0.15&4.2&0.5 &2.3 &0.5&3.2\\ 
			\cline{1-11}
			
			\multirow{1}{*}{\textbf{(e)}}&0.1&2.05&0.3&2.1&0.15&2.5&0.5 &1.5 &0.5&2.2\\ 
			\cline{1-11}
			
			\multirow{1}{*}{\textbf{(f)}}&0.1&2.05&0.3&2.1&0.15&3.8&0.5&1.5&0.5&2.2\\ 
			\cline{1-11}
			
			\multicolumn{11}{|c|}{\textbf{Details of Sharvin's Resistance ($R_S$)}}\\
			\hline
			&&\multicolumn{2}{c|}{weight factor}&\multicolumn{2}{c|}{$ \Delta (meV) $}&\multicolumn{2}{c|}{$ \Gamma (meV) $}  &\multicolumn{2}{c|}{$ Z $}&$ T (K) $\\
			\cline{1-2}\cline{3-4}\cline{5-6}\cline{7-8}\cline{9-11}
			
			\multirow{3}{*}{\textbf{(a)}}&$R_{S1}$&\multicolumn{2}{c|}{0.3}&\multicolumn{2}{c|}{0.7}&\multicolumn{2}{c|}{0.001}  &\multicolumn{2}{c|}{0.4}&2.0\\
			\cline{2-3}\cline{4-5}\cline{5-6}\cline{7-8}\cline{9-11}
			&$R_{S2}$&\multicolumn{2}{c|}{0.35}&\multicolumn{2}{c|}{0.7} &\multicolumn{2}{c|}{0.001}&\multicolumn{2}{c|}{0.45}&2.0\\
			\cline{2-3}\cline{4-5}\cline{5-6}\cline{7-8}\cline{9-11}
			&$R_{S3}$&\multicolumn{2}{c|}{0.3}&\multicolumn{2}{c|}{0.7} &\multicolumn{2}{c|}{0.001}&\multicolumn{2}{c|}{0.55}&2.0\\
			\cline{1-2}\cline{3-4}\cline{5-6}\cline{7-8}\cline{9-11}

			\multirow{3}{*}{\textbf{(b)}}&$R_{S1}$&\multicolumn{2}{c|}{0.35}&\multicolumn{2}{c|}{0.7} &\multicolumn{2}{c|}{0.001}  &\multicolumn{2}{c|}{0.35}&2.0\\
			\cline{2-3}\cline{4-5}\cline{5-6}\cline{7-8}\cline{9-11}
			&$R_{S2}$&\multicolumn{2}{c|}{0.4}&\multicolumn{2}{c|}{0.7} &\multicolumn{2}{c|}{0.001}&\multicolumn{2}{c|}{0.45}&2.0\\
			\cline{2-3}\cline{4-5}\cline{5-6}\cline{7-8}\cline{9-11}
			&$R_{S3}$&\multicolumn{2}{c|}{0.35}&\multicolumn{2}{c|}{0.7} &\multicolumn{2}{c|}{0.001}&\multicolumn{2}{c|}{0.55}&2.0\\
			\cline{1-2}\cline{3-4}\cline{5-6}\cline{7-8}\cline{9-11}
			
			\multirow{3}{*}{\textbf{(c)}}&$R_{S1}$&\multicolumn{2}{c|}{1.0}&\multicolumn{2}{c|}{0.7}&\multicolumn{2}{c|}{0.001}  &\multicolumn{2}{c|}{0.35}&2.0\\
			\cline{2-3}\cline{4-5}\cline{5-6}\cline{7-8}\cline{9-11}
			&$R_{S2}$&\multicolumn{2}{c|}{1.1}&\multicolumn{2}{c|}{0.7} &\multicolumn{2}{c|}{0.001}&\multicolumn{2}{c|}{0.4}&2.0\\
			\cline{2-3}\cline{4-5}\cline{5-6}\cline{7-8}\cline{9-11}
			&$R_{S3}$&\multicolumn{2}{c|}{0.3}&\multicolumn{2}{c|}{0.7} &\multicolumn{2}{c|}{0.001}&\multicolumn{2}{c|}{0.25}&2.0\\
			\cline{1-2}\cline{3-4}\cline{5-6}\cline{7-8}\cline{9-11}
			
			\multirow{3}{*}{\textbf{(d)}}&$R_{S1}$&\multicolumn{2}{c|}{1.0}&\multicolumn{2}{c|}{0.7} &\multicolumn{2}{c|}{0.001}  &\multicolumn{2}{c|}{0.35}&2.0\\
			\cline{2-3}\cline{4-5}\cline{5-6}\cline{7-8}\cline{9-11}
			&$R_{S2}$&\multicolumn{2}{c|}{1.2}&\multicolumn{2}{c|}{0.7} &\multicolumn{2}{c|}{0.001}&\multicolumn{2}{c|}{0.4}&2.0\\
			\cline{2-3}\cline{4-5}\cline{5-6}\cline{7-8}\cline{9-11}
			&$R_{S3}$&\multicolumn{2}{c|}{0.3}&\multicolumn{2}{c|}{0.7} &\multicolumn{2}{c|}{0.001}&\multicolumn{2}{c|}{0.25}&2.0\\
			\cline{1-2}\cline{3-4}\cline{5-6}\cline{7-8}\cline{9-11}
			
			\multirow{3}{*}{\textbf{(e)}}&$R_{S1}$&\multicolumn{2}{c|}{1.0}&\multicolumn{2}{c|}{0.7}&\multicolumn{2}{c|}{0.001}  &\multicolumn{2}{c|}{0.35}&2.0\\
			\cline{2-3}\cline{4-5}\cline{5-6}\cline{7-8}\cline{9-11}
			&$R_{S2}$&\multicolumn{2}{c|}{1.1}&\multicolumn{2}{c|}{0.7} &\multicolumn{2}{c|}{0.001}&\multicolumn{2}{c|}{0.4}&2.0\\
			\cline{2-3}\cline{4-5}\cline{5-6}\cline{7-8}\cline{9-11}
			&$R_{S3}$&\multicolumn{2}{c|}{0.3}&\multicolumn{2}{c|}{0.7} &\multicolumn{2}{c|}{0.001}&\multicolumn{2}{c|}{0.025}&2.0\\
			\cline{1-2}\cline{3-4}\cline{5-6}\cline{7-8}\cline{9-11}
			
			\multirow{3}{*}{\textbf{(f)}}&$R_{S1}$&\multicolumn{2}{c|}{1.0}&\multicolumn{2}{c|}{0.7}&\multicolumn{2}{c|}{0.001}  &\multicolumn{2}{c|}{0.35}&2.0\\
			\cline{2-3}\cline{4-5}\cline{5-6}\cline{7-8}\cline{9-11}
			&$R_{S2}$&\multicolumn{2}{c|}{1.1}&\multicolumn{2}{c|}{0.7} &\multicolumn{2}{c|}{0.001}&\multicolumn{2}{c|}{0.4}&2.0\\
			\cline{2-3}\cline{4-5}\cline{5-6}\cline{7-8}\cline{9-11}
			&$R_{S3}$&\multicolumn{2}{c|}{0.3}&\multicolumn{2}{c|}{0.7} &\multicolumn{2}{c|}{0.001}&\multicolumn{2}{c|}{0.25}&2.0\\
			\cline{1-2}\cline{3-4}\cline{5-6}\cline{7-8}\cline{9-11}
		\end{tabular}
	\end{center}
\end{table}

\begin{table}
	\begin{center}
		\tiny
		\caption{\textbf{The parameters used to generate $V-I$ characteristics in figure 4(a-g) in the manuscript. The values of $R_M$ components and the respective critical current ($I_C (mA)$ ) for the multi-constriction model (shown in figure 1(c)) are listed in ``Details of Maxwell's Resistance ($R_M$)". The parameters to generate $I-V$ characteristics (from BTK theory) corresponding to $R_S$ with weightage factor are listed in ``Details of Sharvin's Resistance ($R_S$)".}}
		\label{tab:table1}
		\begin{tabular}{|p{1.2cm}||p{1.2cm}|p{1.2cm}|p{1.2cm}|p{1.2cm}|p{1.2cm}|p{1.2cm}|p{1.2cm}|p{1.2cm}|p{1.2cm}|p{1.2cm}|}
			
			\hline
			\multicolumn{11}{|c|}{\textbf{Details of Maxwell's Resistance ($R_M$)}}\\
			\hline
			
			&$R_{M1}$ &$I_{C1}$&$R_{M2}$ &$I_{C2}$&$R_{M3}$ &$I_{C3}$ &$R_{M4}$ &$I_{C4}$ &$R_{M5}$&$I_{C5}$\\
			\cline{1-11}
			\multirow{1}{*}{\textbf{(a)}}&0.3&3.5&0.3&3.525&0.4&4.5&0.3&3.45&0.4&3.425\\ 
			\cline{1-11}
			
			\multirow{1}{*}{\textbf{(b)}}&1.0&2.5&1.1&2.525&1.1&3.75&1.05&2.45&1.2&2.425\\ 
			\cline{1-11}
			
			\multirow{1}{*}{\textbf{(c)}}&0.01&0.3&0.05&0.32&0.1&0.45&0.02&0.33&0.03&0.34\\ 
			\cline{1-11}
			
			\multirow{1}{*}{\textbf{(d)}}&0.1&0.2&0.2&0.25&0.3&0.4&0.7&0.3&0.7&0.35\\ 
			\cline{1-11}
			
			\multirow{1}{*}{\textbf{(e)}}&0.01&0.2&0.05&0.25&0.7&1.8&0.02&0.3&0.03&0.35\\ 
			\cline{1-11}
			
			\multirow{1}{*}{\textbf{(f)}}&0.5&1.7&0.5&1.8&0.05&5.5&0.5&0.15&0.5&0.25\\ 
			\cline{1-11}
			
			\multicolumn{11}{|c|}{\textbf{Details of Sharvin's Resistance ($R_S$)}}\\
			\hline
			&&\multicolumn{2}{c|}{weight factor}&\multicolumn{2}{c|}{$ \Delta (meV) $} &\multicolumn{2}{c|}{$ \Gamma (meV) $}  &\multicolumn{2}{c|}{$ Z $}  &$ T (K) $\\
			\cline{1-2}\cline{3-4}\cline{5-6}\cline{7-8}\cline{9-11}
			
			\multirow{3}{*}{\textbf{(a)}}&$R_{S1}$&\multicolumn{2}{c|}{0.3}&\multicolumn{2}{c|}{0.7}&\multicolumn{2}{c|}{0.001}  &\multicolumn{2}{c|}{1.0}&1.6\\
			\cline{2-3}\cline{4-5}\cline{5-6}\cline{7-8}\cline{9-11}
			&$R_{S2}$&\multicolumn{2}{c|}{0.5}&\multicolumn{2}{c|}{0.7} &\multicolumn{2}{c|}{0.001}&\multicolumn{2}{c|}{1.05} &1.6\\
			\cline{2-3}\cline{4-5}\cline{5-6}\cline{7-8}\cline{9-11}
			&$R_{S3}$&\multicolumn{2}{c|}{0.3}&\multicolumn{2}{c|}{0.7} &\multicolumn{2}{c|}{0.001}&\multicolumn{2}{c|}{1.1}&1.6\\
			\cline{1-2}\cline{3-4}\cline{5-6}\cline{7-8}\cline{9-11}

			\multirow{3}{*}{\textbf{(b)}}&$R_{S1}$&\multicolumn{2}{c|}{0.7}&\multicolumn{2}{c|}{0.5} &\multicolumn{2}{c|}{0.001}  &\multicolumn{2}{c|}{0.01}&1.6\\
			\cline{2-3}\cline{4-5}\cline{5-6}\cline{7-8}\cline{9-11}
			&$R_{S2}$&\multicolumn{2}{c|}{0.7}&\multicolumn{2}{c|}{0.5} &\multicolumn{2}{c|}{0.001}&\multicolumn{2}{c|}{0.02}&1.6\\
			\cline{2-3}\cline{4-5}\cline{5-6}\cline{7-8}\cline{9-11}
			&$R_{S3}$&\multicolumn{2}{c|}{0.7}&\multicolumn{2}{c|}{0.5} &\multicolumn{2}{c|}{0.001}&\multicolumn{2}{c|}{0.03} &1.6\\
			\cline{1-2}\cline{3-4}\cline{5-6}\cline{7-8}\cline{9-11}
			
			\multirow{3}{*}{\textbf{(c)}}&$R_{S1}$&\multicolumn{2}{c|}{1.25}&\multicolumn{2}{c|}{1.5}&\multicolumn{2}{c|}{0.001}  &\multicolumn{2}{c|}{0.02}&1.6\\
			\cline{2-3}\cline{4-5}\cline{5-6}\cline{7-8}\cline{9-11}
			&$R_{S2}$&\multicolumn{2}{c|}{1.25}&\multicolumn{2}{c|}{1.5}&\multicolumn{2}{c|}{0.001}&\multicolumn{2}{c|}{0.021}&1.6\\
			\cline{2-3}\cline{4-5}\cline{5-6}\cline{7-8}\cline{9-11}
			&$R_{S3}$&\multicolumn{2}{c|}{1.25}&\multicolumn{2}{c|}{1.5}&\multicolumn{2}{c|}{0.001}&\multicolumn{2}{c|}{0.022}&1.6\\
			\cline{1-2}\cline{3-4}\cline{5-6}\cline{7-8}\cline{9-11}
			
			\multirow{3}{*}{\textbf{(d)}}&$R_{S1}$&\multicolumn{2}{c|}{1.25}&\multicolumn{2}{c|}{1.5}&\multicolumn{2}{c|}{0.001}  &\multicolumn{2}{c|}{0.02}&1.6\\
			\cline{2-3}\cline{4-5}\cline{5-6}\cline{7-8}\cline{9-11}
			&$R_{S2}$&\multicolumn{2}{c|}{1.25}&\multicolumn{2}{c|}{1.5}&\multicolumn{2}{c|}{0.001}&\multicolumn{2}{c|}{0.021}&1.6\\
			\cline{2-3}\cline{4-5}\cline{5-6}\cline{7-8}\cline{9-11}
			&$R_{S3}$&\multicolumn{2}{c|}{1.25}&\multicolumn{2}{c|}{1.5}&\multicolumn{2}{c|}{0.001}&\multicolumn{2}{c|}{0.022}&1.6\\
			\cline{1-2}\cline{3-4}\cline{5-6}\cline{7-8}\cline{9-11}
			
			\multirow{3}{*}{\textbf{(e)}}&$R_{S1}$&\multicolumn{2}{c|}{1.25}&\multicolumn{2}{c|}{1.5}&\multicolumn{2}{c|}{0.001}  &\multicolumn{2}{c|}{0.02}&1.6\\
			\cline{2-3}\cline{4-5}\cline{5-6}\cline{7-8}\cline{9-11}
			&$R_{S2}$&\multicolumn{2}{c|}{1.2}&\multicolumn{2}{c|}{1.5}&\multicolumn{2}{c|}{0.001}&\multicolumn{2}{c|}{0.021}&1.6\\
			\cline{2-3}\cline{4-5}\cline{5-6}\cline{7-8}\cline{9-11}
			&$R_{S3}$&\multicolumn{2}{c|}{1.25}&\multicolumn{2}{c|}{1.5}&\multicolumn{2}{c|}{0.001}&\multicolumn{2}{c|}{0.022}&1.6\\
			\cline{1-2}\cline{3-4}\cline{5-6}\cline{7-8}\cline{9-11}
			
			\multirow{3}{*}{\textbf{(f)}}&$R_{S1}$&\multicolumn{2}{c|}{1.2}&\multicolumn{2}{c|}{1.5}&\multicolumn{2}{c|}{0.001}  &\multicolumn{2}{c|}{0.02}&1.6\\
			\cline{2-3}\cline{4-5}\cline{5-6}\cline{7-8}\cline{9-11}
			&$R_{S2}$&\multicolumn{2}{c|}{1.5}&\multicolumn{2}{c|}{1.5}&\multicolumn{2}{c|}{0.001}&\multicolumn{2}{c|}{0.021}&1.6\\
			\cline{2-3}\cline{4-5}\cline{5-6}\cline{7-8}\cline{9-11}
			&$R_{S3}$&\multicolumn{2}{c|}{1.2}&\multicolumn{2}{c|}{1.5}&\multicolumn{2}{c|}{0.001}&\multicolumn{2}{c|}{0.022}&1.6\\
			\cline{1-2}\cline{3-4}\cline{5-6}\cline{7-8}\cline{9-11}
		\end{tabular}
	\end{center}
\end{table}

\end{document}